\documentclass[twocolumn]{aa} 
\usepackage{graphicx}
\usepackage{txfonts}
\usepackage{natbib}
\bibpunct{(}{)}{;}{a}{}{,} 
\def\kmsmpc{$\rm km\;s^{-1}\;Mpc^{-1}$}
\def\msun{M$_\odot$}
\def\lsun{L$_\odot$}

\begin{document}
\title{Multiband photometric decomposition of nuclear stellar disks}


\author{L. Morelli\inst{1}
        \and
        M. Cesetti\inst{2}
        \and
	E. M. Corsini\inst{1}
        \and
        A. Pizzella\inst{1}
        \and
        E. Dalla Bont\`a\inst{1} 
        \and
        M. Sarzi\inst{3}
        \and
        F. Bertola\inst{1}      
	}

\institute{Dipartimento di Astronomia, Universit\`a di Padova, 
              vicolo dell'Osservatorio~3, I-35122 Padova, Italy\\ 
              \email{lorenzo.morelli@unipd.it} 
           \and INAF-Osservatorio Astronomico di Padova,
                vicolo dell'Osservatorio~2, I-35122 Padova, Italy
           \and Centre for Astrophysics Research, University of Hertfordshire, 
                College Lane, Hatfield, Herts AL10 9AB, United Kingdom}

\date{\today}

\abstract 
{
Small, bright stellar disks with scale lengths of a few tens of
parsec are known to reside in the center of galaxies. They are
believed to have formed in a dissipational process as the end result
of star formation in gas either accreted during a merging (or acquisition)
event or piled up by the secular evolution of a nuclear bar. Only a few
of them have been studied in detail to date.}
{
Using archival Hubble Space Telescope (HST) imaging, we investigate
the photometric parameters of the nuclear stellar disks hosted by
three early-type galaxies in the Virgo cluster, NGC~4458, NGC4478, and
NGC4570, to constrain the process that forms their stars.}
{ 
The central surface brightness, scale length, inclination, and
position angle of the nuclear disks were derived by adopting the
photometric decomposition method introduced by Scorza \& Bender and
assuming the disks to be infinitesimally thin and exponential.}
{
The location, orientation, and size of the nuclear disks is the same
in all the images obtained with the Wide Field Planetary Camera 2 and
Advanced Camera for Surveys and available in the HST Science
Archive. The scale length, inclination, and position angle of each
disk are constant within the errors in the observed $U$, $B$, $V$, and
$I$ passbands, independently of their values and the properties of
the host spheroid.}
{ 
We interpret the absence of color gradients in the stellar population
of the nuclear disks as the signature that star formation
homogeneously occurred along their length. An inside-out
formation scenario is, instead, expected to produce color gradients and
is therefore ruled out.}


\maketitle
%

\section{Introduction}
\label{sec:intro}

The presence of small and bright stellar disks with scale lengths of a
few tens of parsec was first discovered in the nuclear regions of
nearby galaxies about fifteen years ago thanks to the
subarcsec-resolution capabilities of the Hubble Space Telescope
\citep[HST,][]{vdbetal94, kormetal94, emsetal94, laueretal95}.
Afterward, nuclear stellar disks (NSDs) and larger scale embedded
disks were reported in a large number of early-type galaxies as a
results of photometric \citep{laueretal95, caroetal97, ravin01,
  restetal01, tranetal01, ferretal06, sethetal06, lopeetal07,
  kormetal09} and spectroscopic surveys \citep{hallietal01,
  kuntetal06, mcdeetal06} respectively. Given that nuclear disks are
easiest to detect when nearly edge-on \citep{rixwhi90}, the observed
fraction is consistent with NSDs being a common structure in
early-type galaxies \citep{ledoetal10}. At the center of spiral
galaxies NSDs are also present but these are relatively rare in late
types \citep[][]{pizzetal02, falcetal06, peleetal07,balcetal07,moreetal08}.

Photometric parameters of NSDs (i.e., central surface brightness,
scale-length, axial ratio, and position angle) have been derived for
nine galaxies spanning a wide range of Hubble types, from ellipticals
\citep{moreetal04}, to lenticulars \citep{vdbetal94, laueretal95,
  kormetal96a, vdbetal98, scovdb98}, and to early-type spirals
\citep{pizzetal02}.
They are characterized by a smaller scale length ($10-30$ pc) and
higher central face-on surface brightness ($15-19$ mag arcsec$^{-2}$
in the $V$ band) than the embedded stellar disks observed
in ellipticals and lenticulars as well as to the large kpc-scale disks
of lenticulars and spirals \citep{vdb98, pizzetal02, moreetal04}.  
In particular, the embedded stellar disks represent the
intermediary between the sequence of elliptical galaxies and that of
disk galaxies. They have been detected in both disky elliptical
and lenticular galaxies, have scale lengths of $0.1-1$ kpc, and
central face-on surface brightnesses of $18-21$ mag arcsec$^{-2}$ in
the $V$ band. Their angular momentum is parallel to that of their host
spheroid. This demonstrates that the embedded disks are not the result
of accretion or merger events but are likely to be primordial, in the
same sense as the main disks of lenticular and spiral galaxies
\citep{scoben95}.
The smooth variation in the scale parameters from the nuclear to the
embedded and main stellar galactic disks provides observational
support that galaxy properties vary with continuity along a sequence
of increasing disk-to-bulge ratio \citep[see][and references
  therein]{korben96}. Therefore, unveiling the formation scenario of
NSDs may also improve our understanding of galaxy formation and evolution.

In the current picture, NSDs are believed to have formed in a
dissipational process as the end result of star formation in gas
either accreted in a merging (or acquisition) event \citep[the most
  clearcut case is NGC~4698,][]{bertetal99, pizzetal02} or piled up by
the secular evolution of a nuclear bar \citep[e.g.,
  NGC~4570,][]{vdbetal98, scovdb98, vdbems98}.
Each of these scenarios is likely to be correct for some, but not
all, objects. In both of them, the gas is efficiently directed
toward the galaxy center \citep[e.g.,][]{barher96, athaetal05,
  dottetal07, elicetal09, hopqua10}, where it first dissipates and
settles onto an equilibrium plane and then forms into stars.  A
striking example of an on going process of dissipational formation is
provided by NGC~4486A \citep{kormetal05}. The NSD of this
low-luminosity elliptical coexists with its progenitor disk of dust
and gas. The question about the origin of these dynamically-cold
components is also related to their formation epoch, i.e., whether
they built up in the early stages of the galaxy assembly or whether
they formed later.

To address this issue, \citet{vdbetal98}, \citet{vdbems98},
\citet{moreetal04}, and \citet{krajaf04} studied the age and
metallicity content of some NSDs. In some cases, the nuclear disk
was found to be younger than the stellar population of their surrounding
spheroidal component (e.g., NGC~4486A, \citealt{kormetal05}), in other
cases the formation of the disks occurred at the same time as the main
body of the host galaxy (e.g., NGC~4342, \citealt{vdbetal98}).

Characterizing how the star formation proceeded in nuclear disks can
lead to understanding what triggered their formation. For instance, if
stars formed more or less at the same rate throughout the disk from
disk material that accumulated very rapidly at the center, as one
expects to happen during a merging event, we should not find strong
radial gradients in the properties of the stellar
populations. In contrast, color gradients are expected if the disk experiences
an inside-out star formation, like the disks of spiral galaxies
\citep[e.g.,][]{munoetal07}
In this paper, we use HST archival images of the nuclei of three
early-type galaxies in the Virgo cluster, NGC~4458, NGC~4478, and
NGC~4570 to perform a multiband analysis of the photometric parameters
of their NSDs. Our aim is to understand whether the star formation of
the nuclear disk was homogeneous and occurred simultaneously
everywhere along its length.

\section{Sample galaxies}
\label{sec:sample}

We searched the HST Science Archive for all NSDs known in
  literature that had been imaged in at least three filters with the
Wide Field Planetary Camera 2 (WFPC2) and Advanced Camera for Surveys
(ACS) and had known properties of their stellar population.
We found multiband images of the nuclear disks hosted in three
early-type galaxies of the Virgo cluster, NGC~4458, NGC~4478, and NGC
4570 (Table \ref{tab:campione}).

NGC~4458 and NGC~4478 are classified as E0-1 \citep[][hereafter
  RC3]{rc3} and E2 \citep[][hereafter RSA; RC3]{rsa},
respectively. They are low-luminosity ellipticals with a $M_B^0 \sim
-18$ (RC3) at a distance of 12.6 Mpc \citep[][$H_0 = 100$
  \kmsmpc]{tully88}.
Both galaxies host a kinematically-decoupled core \citep{hallietal01,
  emsetal07}. \citet{moreetal04} detected NSDs inside both on the same radial
scale as the core and analyzed their structure and stellar content. Their
scale length and face-on central surface brightness (Table
\ref{tab:disk_parameters}) were derived from WFPC2/F814W images and
they fit in the relation for NSDs. For NGC~4458, these parameters are
typical of NSDs, while for NGC~4478 they are in-between those of NSDs
and the disks of disky ellipticals. The mass of the two disks is of
few $10^7$ \msun\/.
The central disk of NGC~4458 is counter-rotating. It has similar
properties to the decoupled cores of bright ellipticals
\citep[see][for a review]{Bercors99}. According to \citet{moreetal04},
it is as old as ($t \simeq 15$ Gyr) and richer in metals ($[Z/{\rm H}]
\simeq 0.2$) than the rest of the galaxy, with a high and approximately
constant overabundance ($[\alpha/{\rm Fe}] \simeq 0.3$).
The NSD of NGC~4478 is co-rotating with respect to the main body of
its host galaxy, but it is younger ($t \simeq 6$ Gyr), richer in
metals ($[Z/{\rm H}] \simeq 0.4$) and less overabundant
($[\alpha/{\rm Fe}] \simeq 0.2$) than the outer regions. The nearly
solar $\alpha$-element abundance indicates a prolonged star formation
history, typical of an undisturbed disk-like, gas-rich, and possibly
pre-enriched structure.
\citet{moreetal04} used the models of single stellar populations with
$\alpha$/Fe overabundances developed by \citet{thmabe03} to derive the properties
of the nuclear stellar populations. They analyzed the absorption
line-strength indices measured along the major and minor axes of both
galaxies. The masses of the two NSDs were estimated by adopting the
mass-to-light ratios calculated with the models of \citet{mara98} from
the available ages and metallicities.
The presence of cold disks at the centers of NGC~4458 and NGC~4478
argues for a formation scenario based on the dissipational collapse of
a gas-rich object. An external origin of the gas is claimed to account
for the counter-rotation of NGC~4478 \citep{moreetal04}.

NGC~4570 is a edge-on disk galaxy classified as S0 by RSA and RC3. It
has $M_B^0 \sim -19$ (RC3) at a distance of 12.6 Mpc \citep{tully88}.
Its nuclear disk was first studied in detail by \citet{vdbetal98}. It
resides within the inner cutoff of the main stellar disk, whose light
contribution to the galaxy surface brightness is negligible within the
inner $2''$. The inner truncation of the main disk is a general
feature in multi-disk systems hosting a nuclear \citep{vdbetal98,
  pizzetal02} and/or an embedded disk \citep[e.g.,][]{scoben95,
  seisco96, emsetal96}. The photometric parameters (Table
\ref{tab:disk_parameters}) were derived by analyzing a WFPC2/F555W
image and are typical of NSDs \citep{scovdb98}.
The NSD of NGC~4570 has a mass of $8 \times 10^7$ \msun\ (at the
adopted distance) as derived from dynamical modelling by
\citet{vdbems98}. It has a different stellar population than the rest
of the bulge, consisting of relatively old ($t \simeq 8$ Gyr) and
metal-rich ($[{\rm Fe}/{\rm H}] \simeq 0.35$) stars
\citep{vdbetal98}. These results were derived by comparing the colors
measured in WFPC2 images and the line-strength indices measured in
Faint Object Spectrograph spectra with the models of single stellar
populations by \citet{Worth94}.

In addiction, NGC~4570 exhibits two nuclear edge-on rings. According to
dynamical modelling by \citet{vdbems98} based on the observed surface
brightness and stellar kinematics, the position of the rings (at about
$2''$ and $4''$ from the center, respectively) is consistent with the
location of the inner Lindblad and ultra-harmonic resonances of a
rapidly-rotating nuclear bar of about 500 pc. The corotation radius is
close to the inner cut-off of the main disk at about $7''$ from the
center.
The morphology, dynamics, star formation, and chemical enrichment of
the nuclear region strongly favor a formation scenario of the NSD
driven by the secular evolution. The nuclear disk appears to be the
end result of the star formation that occurred in the gas that was
funneled to the center by the nuclear bar \citep{vdbems98,
  krajaf04}. The growth of a central mass concentration produced by
the gas inflow altered the shape of the orbits sustaining the bar. As
a consequence, the nuclear bar weakened and dissolved \citep[see][and
  references therein]{athaetal05}.

\begin{table*}[t]
\caption{Details about the HST archival images of the sample galaxies.}
\label{tab:campione}
\begin{center}
\begin{tabular}{l l c c c r c l}
\hline
\hline
\noalign{\smallskip}
\multicolumn{1}{c}{Galaxy} & 
\multicolumn{1}{c}{Camera} & 
\multicolumn{1}{c}{Filter} & 
\multicolumn{1}{c}{Obs. Date} & 
\multicolumn{1}{c}{No. Exp.} & 
\multicolumn{1}{c}{Exp. Time} & 
\multicolumn{1}{c}{Prog. Id.} & 
\multicolumn{1}{c}{P.I.}\\
\noalign{\smallskip}
\multicolumn{1}{c}{} & 
\multicolumn{1}{c}{} & 
\multicolumn{1}{c}{} & 
\multicolumn{1}{c}{} & 
\multicolumn{1}{c}{} & 
\multicolumn{1}{c}{[s]} & 
\multicolumn{1}{c}{} & 
\multicolumn{1}{c}{}\\
\noalign{\smallskip}
\multicolumn{1}{c}{(1)} &
\multicolumn{1}{c}{(2)} &
\multicolumn{1}{c}{(3)} &
\multicolumn{1}{c}{(4)} &
\multicolumn{1}{c}{(5)} &
\multicolumn{1}{c}{(6)} &
\multicolumn{1}{c}{(7)} &
\multicolumn{1}{c}{(8)} \\
\noalign{\smallskip}
\hline		    
\noalign{\smallskip}        	                		 
NGC 4458 & ACS/WFC  & F475W & 28 Dec 2003 & 2 &  750 & 9401 & P. C\^ot\'e  \\
         & WFPC2/PC & F555W & 18 Dec 1998 & 3 & 1340 & 6587 & D. O. Richstone\\
         & WFPC2/PC & F814W & 12 Feb 1995 & 6 & 1120 & 5512 & S. M. Faber \\
NGC 4478 & ACS/HRC  & F330W & 28 jul 2005 & 4 & 1912 & 10435& R. W. O'connell\\
         & ACS/WFC  & F475W & 09 Jul 2003 & 2 &  750 & 9401 & P. C\^ot\'e  \\
         & WFPC2/PC & F555W & 07 Jul 1997 & 4 & 1600 & 6587 & D. O. Richstone\\
         & WFPC2/PC & F814W & 07 Jul 1997 & 3 & 1600 & 6587 & D. O. Richstone\\
NGC 4570 & WFPC2/PC & F336W & 19 Apr 1996 & 5 & 3100 & 6107 & W. Jaffe\\
         & ACS/WFC  & F475W & 13 Jul 2003 & 2 &  750 & 9401 & P. C\^ot\'e \\ 
	 & WFPC2/PC & F555W & 19 Apr 1996 & 2 &  400 & 6107 & W. Jaffe\\
         & WFPC2/PC & F814W & 19 Apr 1996 & 2 &  460 & 6107 & W. Jaffe\\        
\noalign{\smallskip}
\hline				    	    			 
\end{tabular}
\end{center}
\begin{minipage}{18.5cm}
\begin{small}
NOTES: Col.(1): Galaxy name. Col.(2): Camera name. Col.(3): Filter
name. Col.(4): Observation date. Col.(5): Number of
exposures. Col.(6): Total exposure time. Col.(7): HST proposal number.
Col.(8): Name of the principal investigator. 
\end{small}
\end{minipage}
\end{table*}

\section{Data reduction}
\label{sec:reduction}

The WFPC2 and ACS multiband images of NGC~4458, NGC~4478, and NGC~4570
were retrieved from the HST Science Data Archive. Their list and
observing log are given in Table~\ref{tab:campione}.

The WFPC2 images were obtained by centering the galaxy nucleus on the
higher resolution Planetary Camera (PC). The PC detector is a Loral
CCD with $800 \times 800$ pixels. The pixel size is $15 \times 15$
$\mu$m$^2$. The plate scale of $0\farcs046$ pixel$^{-1}$ yields a
field of view of about $36''\times 36''$.  To help in identifying and
correcting cosmic-ray events, different exposures were taken with each
filter.
The ACS images were taken with both the Wide Field Channel (WFC) and
the High Resolution Channel (HRC). WFC consists of two SITe CCDs with
$2048 \times 4096$ pixels each of size $15 \times 15$
$\mu$m$^2$. The plate scale is $0\farcs049$ pixel$^{-1}$ and the field
of view of the combined detectors covers an approximately square area
of about $202'' \times 202''$.
HRC detector is a SITe CCDs with $1024 \times 1024$ pixels each of
size $21 \times 21$ $\mu$m$^2$ with a plate scale of $0\farcs028
\times 0\farcs025$ pixel$^{-1}$. Because of the large, but well
characterized, geometric distortion affecting the ACS, the $29''
\times 26''$ HRC field of view projects onto the plane of the sky as a
rhomboid with $x$ and $y$ axes forming a $84\fdg2$ angle.
For both WFPC2 and ACS images, the telescope was guiding in fine lock,
giving a typical rms tracking error per exposure of $0\farcs005$.
 
The WFPC2 images were calibrated using the CalWFPC reduction pipeline
in IRAF\footnote{Imaging Reduction and Analysis Facilities (IRAF) is
  distributed by the National Optical Astronomy Observatories which
  are operated by the Association of Universities for Research in
  Astronomy (AURA) under cooperative agreement with the National
  Science Foundation.} maintained by the Space Telescope Science
Institute. Reduction steps include bias subtraction, dark current
subtraction, and flat-fielding, as described in detail in the WFPC2
instrument and data handbooks \citep{baggWFPC,McMaster2008}.
Subsequent analysis was performed using IRAF standard tasks.
The bad pixels were corrected by means of a linear one-dimensional
interpolation using the data quality files and the WFIXUP task. For
each galaxy, the alignment of the images obtained with the same filter
was checked by comparing the centroids of stars in the field of view.
The images were aligned to an accuracy of a few hundredths of a pixel
using IMSHIFT and knowledge of the offsets. They were then combined
with IMCOMBINE.
We verified that the alignment and combination did not introduce a
significant blurring of the data. To this aim, we measured the FWHM of
the Gaussian fitting to the field stars in the original and combined
frames. We found that they did not change to within a few percent.
In combining the images, pixels deviating by more than three times the
local standard deviation -- calculated from the combined effect of
Poisson and read-out noise -- were flagged as cosmic rays and
rejected.  The residual cosmic rays and bad pixels were corrected by
manually editing the resulting image with IMEDIT.

The ACS images were calibrated using the CalACS reduction pipeline.
Reduction steps include bias subtraction, dark current subtraction,
flat-fielding correction, and correction for geometric distortion with
MULTIDRIZZLE task of IRAF as described in detail in ACS instrument and
data handbooks \citep{pav04a, pav04b}.  The images obtained in the
same filter were aligned by comparing the centroids of stars in the
field of view and then combined, rejecting cosmic rays in the process.
Residual cosmic ray events and hot pixels were removed using the
LACOS\_IMA procedure \citep{vand01}.

The sky level in WFPC2 images was determined from apparently empty
regions in the Wide Field chips and subtracted from the PC frame after
appropriate scaling. In ACS images, it was determined from regions free
of sources at the edge of the field of view and then subtracted.
The ACS/F330W and WFPC2/F336W passbands approximate Johnson-Cousins
$U$ band. The ACS/F475W, WFPC2/F555W and WFPC2/F814W passbands are
similar to Johnson-Cousins $B$, $V$, and $I$ bands, respectively.
The flux calibration to the Vega magnitude system was performed
following \citet{holteta95b} for the WFPC2 images, and
\citet{sirietal05} for the ACS images.

\begin{figure*}
\begin{center}
\includegraphics[width=17cm]{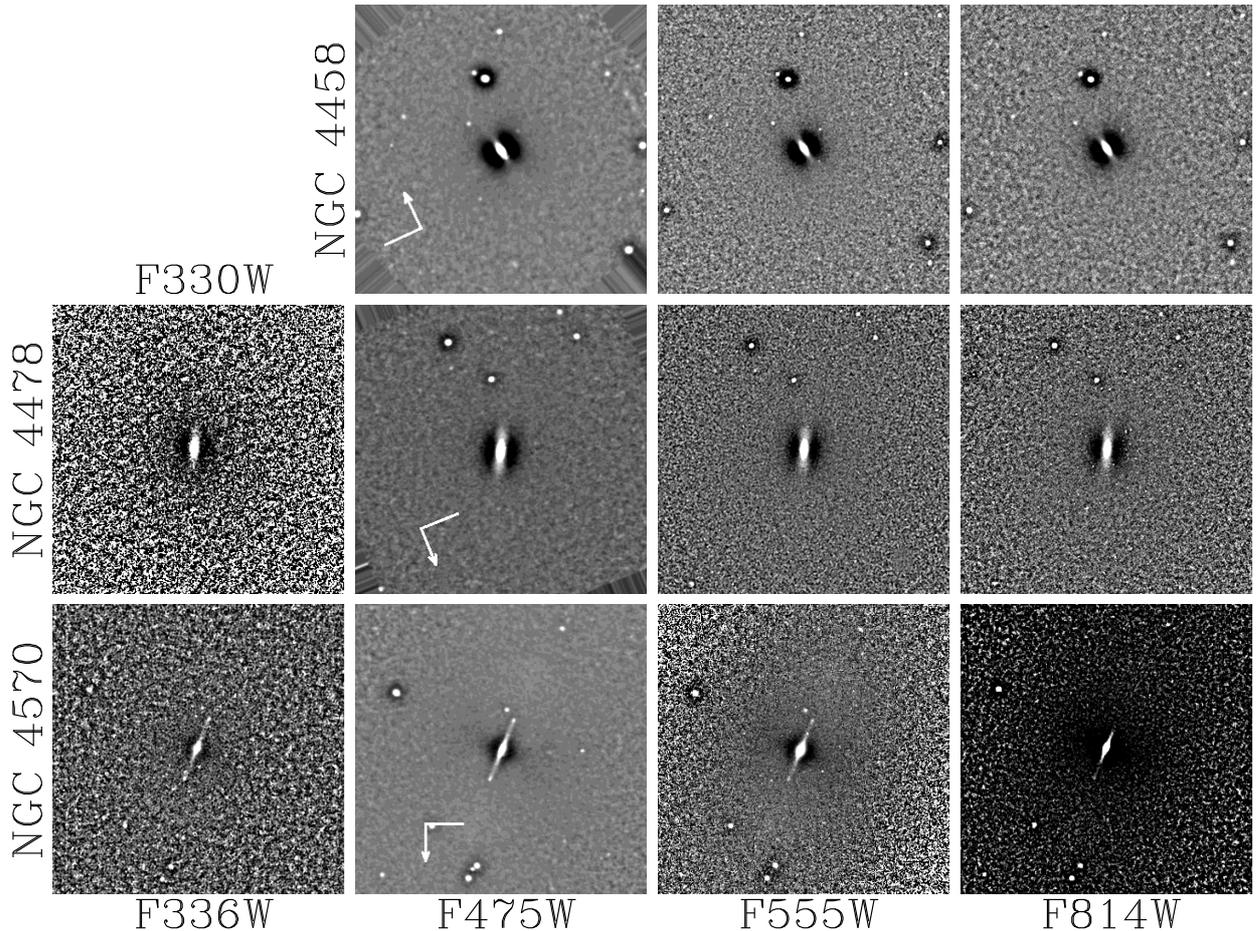}
\caption[]{Unsharp masking of the HST archival images of NGC~4458
  (ACS/F475W, WFPC2/F555W, and WFPC2/F814W; top panels), NGC~4478
  (ACS/F330W, ACS/F475W, WFPC2/F555W, and WFPC2/F814W; middle panels),
  and NGC~4570 (WFPC2/F336W, ACS/F475W, WFPC2/F555W, and WFPC2/F814W;
  bottom panels).  The orientation is specified by the arrow
  indicating north and the segment indicating east in the lower-left
  corner of the ACS/F475W image.  The size of the plotted region is
  about $15'' \times 15''$.}
\label{fig:unsharp}
\end{center}
\end{figure*}

\section{Unsharp images of nuclear disks}
\label{sec:detection}

\begin{figure}
\includegraphics[height=9cm,angle=90]{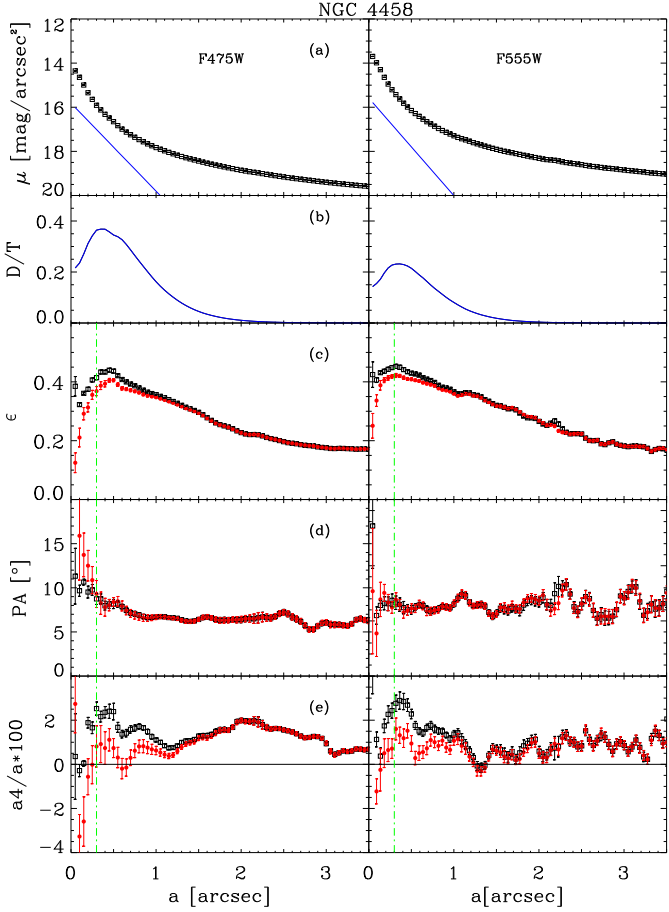}
\caption[]{Isophotal parameters of the nuclear region of NGC~4458 as a
  function of the isophotal semi-major axis. We show the results based
  on the analysis of the surface-brightness distribution measured in
  the ACS/F475W ({\em left panels}) and WFPC2/F555W ({\em right
    panels}) images, respectively. {\em (a)} Surface-brightness radial
  profiles of the galaxy after deconvolution ({\em open black
    squares}) and of the nuclear disk ({\em solid blue line}). {\em
    (b)} Radial profile of the fraction of total luminosity
  contributed by the nuclear disk. Radial profiles of the galaxy
  ellipticity {\em (c)}, position angle {\it (d)\/}, and fourth cosine
  Fourier coefficient {\em (e)} before ({\em open black squares}) and
  after ({\em filled red circles}) the subtraction of the best-fitting
  model for the nuclear disk. The results of the photometric
  decomposition of the WFPC2/F814W image are in \citet{moreetal04}.
  The vertical dash-dotted line indicates the radius inside which the
  deconvolution algorithm is not able to recover the geometric
  parameters of the isophotes.}
\label{fig:plot4458}
\end{figure}

\begin{figure}
\includegraphics[height=9cm,angle=90]{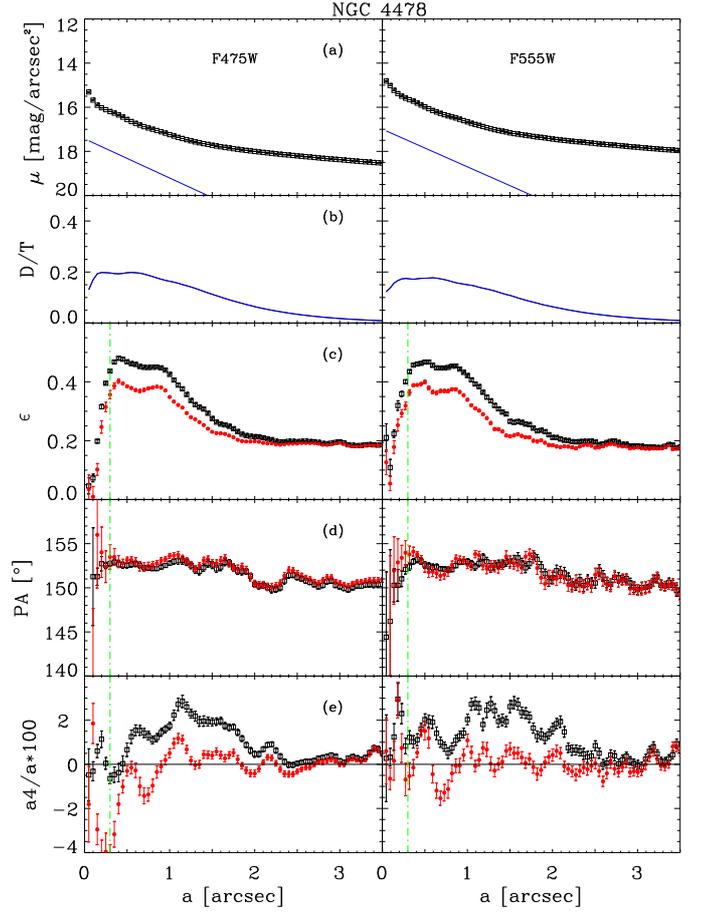}
\caption[]{Same as in Fig.~\ref{fig:plot4458}, but for NGC~4478. The
  results of the photometric decomposition of the WFPC2/F814W image
  are in \citet{moreetal04}.}
\label{fig:plot4478}
\end{figure}

\begin{figure*}
\includegraphics[height=18cm,angle=90]{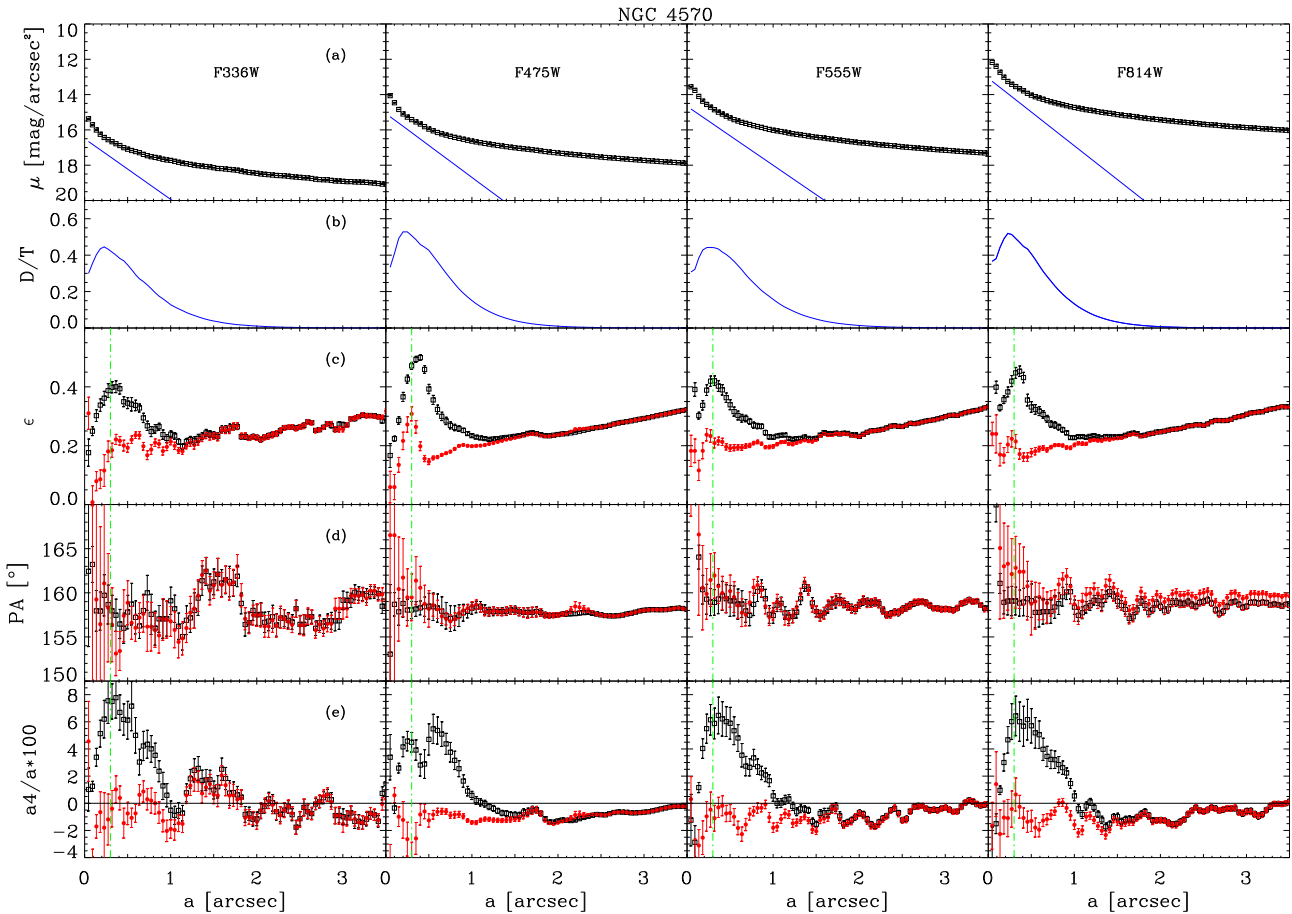}
\caption[]{Same as in Fig.~\ref{fig:plot4458}, but for the
  WFPC2/F336W, ACS/F475W, WFPC2/F555W, and WFPC2/F814W images of
  NGC~4570.}
\label{fig:plot4570}
\end{figure*}

To first gauge the structure and extent of the NSDs in all the
available images of the sample galaxies, we constructed the
unsharp-masked image of each frame using an identical procedure
described by \citet{pizzetal02}. The ACS/HRC image was resampled to
match the same spatial scale as the WFPC2/PC images. Each image was
divided by itself after convolution by a circular Gaussian of width
$\sigma=6$ pixels, corresponding to $0\farcs28$ and $0\farcs29$ in the
WFPC2/PC and ACS/WFC images, respectively (Fig.~\ref{fig:unsharp}).
This procedure enhanced any surface-brightness fluctuation and
non-circular structure extending over a spatial region comparable to
the $\sigma$ of the smoothing Gaussian. 
Each galaxy nucleus clearly contains a highly elongated
structure in all the images. The location, orientation, and size of
each structure is similar in all the observed passbands. These nuclear
structures are associated with a central increase in ellipticity
measured by performing an isophotal analysis using the IRAF task
ELLIPSE (Figs. \ref{fig:plot4458} -- \ref{fig:plot4570}; see also
\citealt{scovdb98}, \citealt{moreetal04}). They are not artifacts of
the unsharp-masking procedure as discussed by \citet{moreetal04} for
NGC~4458 and NGC~4478 and by \citet{scovdb98} for NGC~4570.

The ACS/F475W unsharped-masked image of NGC~4570 exhibits two symmetric
bright spots located along the major axis of the NSD at about $2''$
from the center. They were identified by \citet{vdbems98} with
the signature of the nuclear ring located at the inner Lindblad
resonance in the equatorial plane of the galaxy.  This ring is seen
edge-on. The two spots are barely visible in the WFPC2/F336W and
WFPC2/F555W images and almost missing in the WFPC2/F814W
image. According to \citet{vdbems98}, the blue colors of these
features correspond to a rather young stellar population with an age
of less than 2 Gyr and a metallicity close to solar. The width of the
filtering Gaussian we adopted in the unsharp-masking process is
effective in detecting the NSD, but does not enhance the structure of
the ring located at the ultra-harmonic resonance nor the inner
cutoff to the main disk (for this purpose, see the unsharp-masked
images obtained with different filtering Gaussians that
\citealt{vdbems98} displayed in their Fig.~1).

In \citet{moreetal04}, we derived isophotal profiles from the
analysis of the WFPC2/F814W images of NGC~4458 and NGC~4478. Here, we
present the isophotal profiles derived from the remaining images of
the sample galaxies adopting the same method.
We first masked foreground stars and then fitted ellipses to the
isophotes. We allowed the centers of the ellipses to vary, to test
whether the galaxies were disturbed. Within the errors of the fits, we
found no evidence of variations in the fitted center. The ellipse
fitting was repeated with the ellipse centers fixed. 
The signal-to-noise ratio of the ACS/F330W image of NGC~4478 did not
allow us to measure a reliable radial profile of the $a_4$
coefficient. This is necessary to derive the NSD parameters, therefore
we did not consider this image in the subsequent analysis.
The resulting
azimuthally averaged surface brightness, ellipticity, position angle,
and the fourth cosine Fourier coefficient ($a_4$) profiles of
NGC~4458, NGC~4478, and NGC~4570 are presented in
Figs. \ref{fig:plot4458}, \ref{fig:plot4478}, and \ref{fig:plot4570},
respectively. 
In the innermost about $1''$ of NGC~4458 and NGC~4570 and about $2''$
of NGC~4478, we measured positive values of the $a_4$ Fourier
coefficients, which describe the disky deviation of the isophotes from
pure ellipses \citep{jed87}. These photometric features confirm the
presence of a NSD \citep[see also][]{kormetal09}.

\section{Structural parameters of nuclear disks}
\label{sec:decomposition}

After establishing the existence of NSDs in all passbands, we
derived their photometric parameters using the method of
\citet{scoben95} as implemented by \citet{moreetal04}. This method is
based on the assumption that the isophotal diskiness is the result of
the superimposition of a spheroidal component (which is either a host
elliptical galaxy or a bulge component) and an inclined exponential
disk. The two components are assumed to have both perfectly elliptical
isophotes with constant but different ellipticities.

\begin{figure*}
\begin{center}
\includegraphics[height=10cm,angle=0]{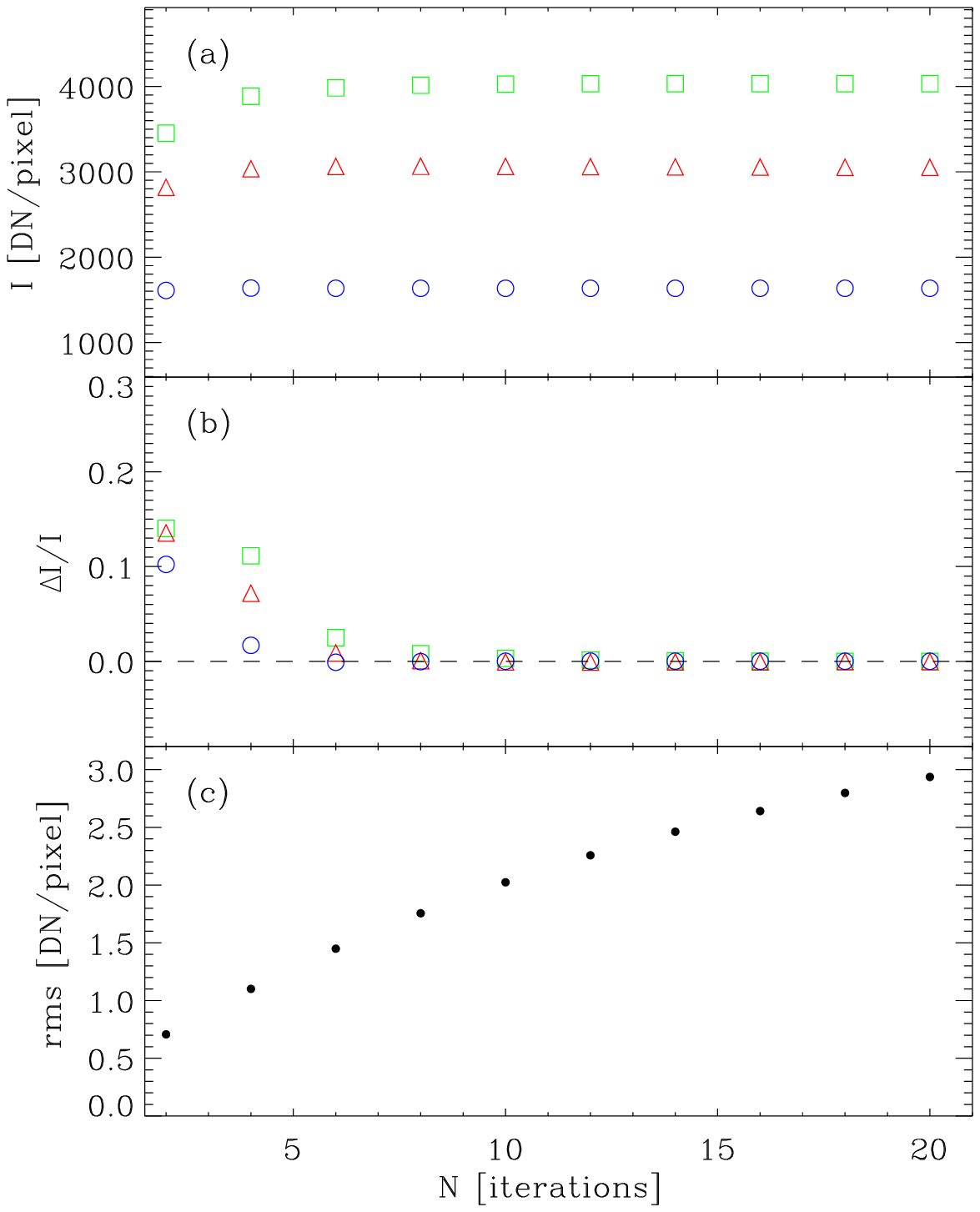}
\includegraphics[height=10cm,angle=0]{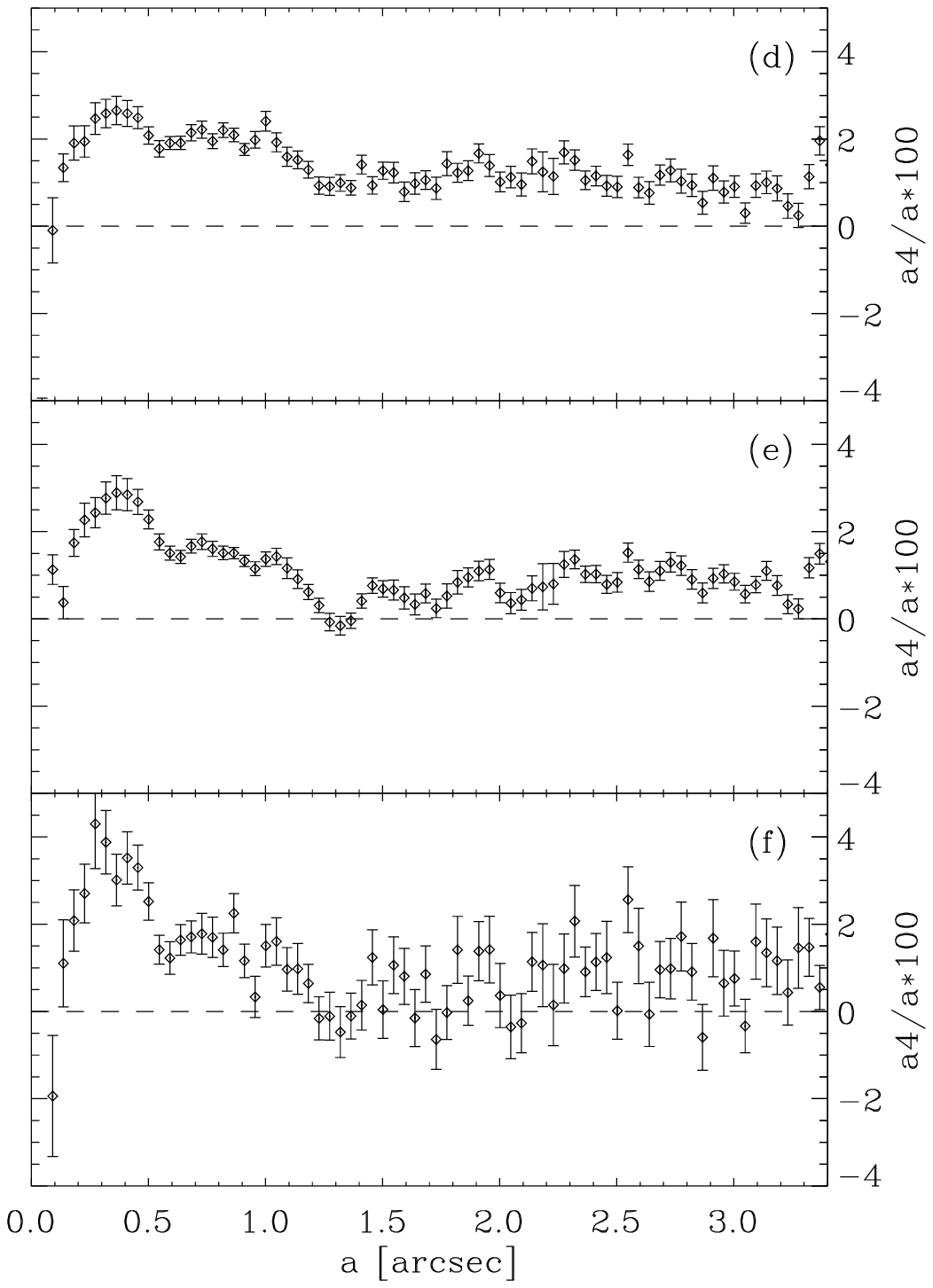}
 \caption[]{ Deconvolution of the WFPC2/F555W image of NGC~4458
   with the Richardson-Lucy algorithm.
  The left panels show {\em (a)} the average surface brightness per
  pixel in the galaxy nucleus, {\em (b)} its relative increase, and
  {\em (c)} the noise level of the image plotted as a function of the
  number of the iterations of the deconvolution process.
  The different symbols refer to the values measured in circular
  apertures with radii of $2$ ({\em green squares}), $3$ ({\em red
    triangles}), and $6$ ({\em blue circles}) pixels centered on the
  galaxy nucleus, respectively.
  The relative increase in surface brightness at iteration 1 is
  calculated with respect to the original undeconvolved image.
  The noise level ({\em filled circles}) of the image was measured
  after each iteration in the same empty regions at the edges of the
  field of view.
  The right panels show the radial profiles of the fourth cosine
  Fourier coefficients measured in the {\em (d)} original image, and
  in the deconvolved image after {\em (e)} 6 and {\em (f)} 20
  iterations.}
\label{fig:deconv}
\end{center}
\end{figure*}

To properly derive the photometric parameters of the nuclear disks, it
is important to account for the HST point-spread function (PSF). The
adopted PSF model was calculated with the TINYTIM package by taking
into account the instrumental setup and position of the nuclear disk
on the given image \citep{krihoo99}.  To restore the images from the
effects of the PSF, they were deconvolved with the Richardson-Lucy
method \citep{rich72, lucy74} by applying the IRAF task LUCY for a
number of iterations between 3 and 6. 
A larger number of iterations was not found to improve the restoration
of the geometrical parameters of the isophotes as discussed in detail
by \citet{mich96} with extensive experiments on the ground and HST-based
images. The image noise was, instead, amplified and errors in the
ellipticity and $a_4$ Fourier coefficients are enlarged.
This effect is illustrated in Fig. \ref{fig:deconv} for the
WFPC2/F555W image of NGC~4458. The average surface brightness in the
galaxy nucleus and the image noise are plotted as a function of the
number of iterations performed with the deconvolution algorithm.
The average surface brightness of the galaxy was measured in circular
apertures of radii 2, 3, and 6 pixels, respectively and centered on
the nucleus. They correspond approximately to 3, 5, and 10 times the
standard deviation of the best-fit Gaussian model of the PSF. The
noise was estimated by measuring the standard deviation in the surface
brightness level for a number of empty regions at the edges of the
field of view.
After 6 iterations, a negligible rise ($<2\%$) in the surface
brightness is observed even in the innermost aperture, which is the
most sensitive to the deconvolution effect.
In contrast, the noise level increases continuously at each
iteration. This increases the uncertainties in the
$a_4$ Fourier coefficients, which are critical for recovering the
structural parameters of the NSD with the method of photometric
decomposition by \citet{scoben95}.
The radial profiles of the $a_4$ Fourier coefficients measured in the
original WFPC2/F555W image of NGC~4458 as well as in the images
resulting after 6 and 20 deconvolution iterations are shown in
Fig. \ref{fig:deconv}. The redistribution of the galaxy light caused
by the deconvolution is characterized by an enhancement of the galaxy
diskiness in the inner $1''$. The $a_4$ Fourier coefficients measured
after 6 and 20 iterations have the same radial trend and are
consistent within the errors. However, the $a_4$ errorbars increased
by a factor of 3 after 20 iterations. We found that after 40
Richardson-Lucy iterations the scatter and errorbars of the $a_4$
Fourier coefficients are sufficiently large to blur the peak in the
$a_4$ radial profile and significantly reduce the radial range along
which the diskiness is observed. This prevents us obtaining accurate
and reliable values of the NSD parameters.

The rms tracking error is negligible with respect to the PSF FWHM
($0\farcs07$), thus it does not affect the image quality and the
measurement of the structural parameters of the NSDs. To this aim, no
correction for telescope jitter was necessary.

\begin{table*}[t]
\caption{Photometric parameters of the stellar nuclear disks.}
\label{tab:disk_parameters}
\begin{center}
\begin{tabular}{l c c c c c c c c}
\hline
\hline
\noalign{\smallskip}
\multicolumn{1}{c}{Galaxy} &
\multicolumn{1}{c}{Filter} &
\multicolumn{1}{c}{$\mu_0$} &
\multicolumn{2}{c}{$h$} &
\multicolumn{1}{c}{$i$} &
\multicolumn{1}{c}{PA} &
\multicolumn{1}{c}{$L$}&
\multicolumn{1}{c}{Ref.}\\
\noalign{\smallskip}
\multicolumn{1}{c}{} &
\multicolumn{1}{c}{} &
\multicolumn{1}{c}{[mag~arcsec$^{-2}$]} &
\multicolumn{1}{c}{[$''$]} &
\multicolumn{1}{c}{[pc]} &
\multicolumn{1}{c}{[$^\circ$]} &
\multicolumn{1}{c}{[$^\circ$]} &
\multicolumn{1}{c}{[$10^6$ \lsun]} &
\multicolumn{1}{c}{}\\
\noalign{\smallskip}
\multicolumn{1}{c}{(1)} &
\multicolumn{1}{c}{(2)} &
\multicolumn{1}{c}{(3)} &
\multicolumn{2}{c}{(4)} &
\multicolumn{1}{c}{(5)} &
\multicolumn{1}{c}{(6)} &
\multicolumn{1}{c}{(7)} &
\multicolumn{1}{c}{(8)} \\
\noalign{\smallskip}
\hline
\noalign{\smallskip}
NGC 4458& F475W & 16.0$^{+0.3}_{-1.0}$ & 0.24$^{+0.30}_{-0.07}$ & 15$^{+18}_{-4 }$  & 84$^{+ 6}_{-13}$  & 7$\pm 1$  &  2.7$^{+3.3}_{-0.5}$      & 1 \\
        & F555W & 15.8$^{+0.2}_{-0.9}$ & 0.24$^{+0.02}_{-0.04}$ & 15$^{+ 1}_{-2 }$  & 82$^{+ 8}_{-6}$   & 7$\pm 1$  &  3.3$^{+3.5}_{-0.5}$      & 1 \\
        & F814W & 14.3$^{+0.7}_{-0.7}$ & 0.18$^{+0.99}_{-0.04}$ & 11$^{+ 6}_{-2 }$  & 83$^{+7}_{-9 }$   & 7$\pm 1$  &  3.0$^{+3.0}_{-2.7}$      & 2 \\ 
NGC 4478& F475W & 17.5$^{+1.4}_{-0.3}$ & 0.55$^{+0.54}_{-0.34}$ & 34$^{+32}_{-21}$  & 78$^{+12}_{-12}$  & 153$\pm 1$  &  6.8$^{+0.9}_{-2.1}$      & 1 \\
	& F555W & 17.1$^{+1.3}_{-0.5}$ & 0.63$^{+0.95}_{-0.25}$ & 38$^{+58}_{-15}$  & 78$^{+12}_{-11}$  & 153$\pm 1$  & 10.5$^{+6.2}_{-3.6}$      & 1 \\
	& F814W & 15.8$^{+1.5}_{-0.8}$ & 0.66$^{+0.54}_{-0.30}$ & 40$^{+33}_{-19}$  & 80$^{+10}_{-10}$  & 153$\pm 1$  & 17.0$^{+5.6}_{-6.3}$      & 2 \\
NGC 4570& F336W & 16.6$^{+0.7}_{-1.0}$ & 0.31$^{+0.32}_{-0.04}$ & 19$^{+20}_{-2 }$  & 83$^{+7}_{-6}$    & 158$\pm 1$  &  4.5$^{+7.9}_{-2.5}$      & 1 \\
        & F475W & 15.3$^{+0.3}_{-0.9}$ & 0.27$^{+0.32}_{-0.07}$ & 16$^{+20}_{-4 }$  & 78$^{+12}_{-7}$   & 158$\pm 1$  & 13.7$^{+6.9}_{-1.4}$      & 1 \\
	& F555W & 14.8$^{+0.7}_{-1.0}$ & 0.32$^{+0.40}_{-0.05}$ & 19$^{+25}_{-3 }$  & 83$^{+7}_{-3}$    & 158$\pm 1$ & 12.9$^{+16.0}_{-5.0}$     & 1 \\
	& F555W & 14.8~~~~~          & ...                  & 23~~~~          & ...~            & ... & 19~~~~~~~~~             & 3 \\
	& F814W & 13.2$^{+0.9}_{-0.9}$ & 0.28$^{+0.35}_{-0.08}$ & 17$^{+22}_{-5 }$  & 81$^{+ 9}_{- 6 }$ & 158$\pm 1$  & 22.4$^{+22.0}_{-10.0}$    & 1 \\
\noalign{\smallskip}
\hline				    	    			 
\end{tabular}
\end{center}
\begin{minipage}{18.5cm}
\begin{small}
NOTES: Col.(1): Galaxy name. Col.(2): Filter name. Col.(3): Disk
central surface-brightness in the Vega magnitude system. Col.(4): Disk
scale length. A distance of 12.6 Mpc was adopted for all the sample
galaxies. Col.(5): Disk inclination Col.(6) Disk position
angle. Col.(7): Disk luminosity. Col.(8): Reference paper for the
photometric decomposition.  1 = this work, 2 = \citet{moreetal04}, 3 =
\citet{scovdb98}.
\end{small}
\end{minipage}
\end{table*}

The \citet{scoben95} method consists of the iterative subtraction of a
thin disk model characterized by an exponential surface-brightness
profile of central surface brightness $\mu_0$, scale-length $h$, axial
ratio $b/a$, and position angle PA. The disk inclination was
calculated to be $i=\arccos{(b/a)}$.  The position angle of each
nuclear disk was found to be constant within $1^\circ$ for all the
images of the same galaxy (Table~\ref{tab:disk_parameters}). Adopting
this value, the remaining disk parameters are adjusted until the
departures from perfect ellipses are minimized (i.e.,
$a_{4}\approx0$).

\citet{ferretal06} measured in HST/ACS images the isophotal parameters
of sample of early-type members of the Virgo cluster (including
NGC~4458, NGC~4478, and NGC~4570). Inside $r < 0.1 r_{\rm e}$, there
is a trend for the mean $a_4$ to increase as galaxies become
fainter. On large scales ($0.1 r_{\rm e} < r < r_{\rm e}$), the
isophotes are consistent with having a constant shape. In this radial
range, most (but not all) of the observed galaxies exhibit almost
elliptical isophotes. These results agree with the previous
findings based on HST photometry (\citealt{vdbetal94};
\citealt{laueretal95}; \citealt{veretal99}; \citealt{restetal01}; but
see also \citealt{kormetal09}).
We therefore decided to adjust the NSD parameters until the isophotes
of the host spheroid had the same shape observed outside the inner
region where $a_4>0$. We adopted $a_{4} \simeq 0.01$ for NGC~4458,
$a_{4} \simeq 0$ for NGC~4478, and $a_{4}\simeq -0.01$ for NGC~4570,
respectively.

For each disk model, the disk-free image of the galaxy was obtained by
subtracting the disk model from the galaxy image. The isophotal
analysis of the disk-free image was performed using the IRAF task
ELLIPSE.  The photometric decomposition was performed independently
for each image. The best-fit values of $\mu_0$, $h$, and $b/a$ and
their $68\%$ confidence-level errors were derived in a similar way to
those in \citet{moreetal04} and are listed in
Tab.~\ref{tab:disk_parameters}.
The comparison between the isophotal parameters of the galaxies
measured before and after the subtraction of the best-fit model of
their nuclear disk are shown in Figs. \ref{fig:plot4458} --
\ref{fig:plot4570}.
The surface-brightness radial profiles of the nuclear disks and the
radial profile of the light fraction $D/T$ that they contribute to the
galaxy luminosity are also plotted.

\begin{figure}
\includegraphics[width=9.4cm]{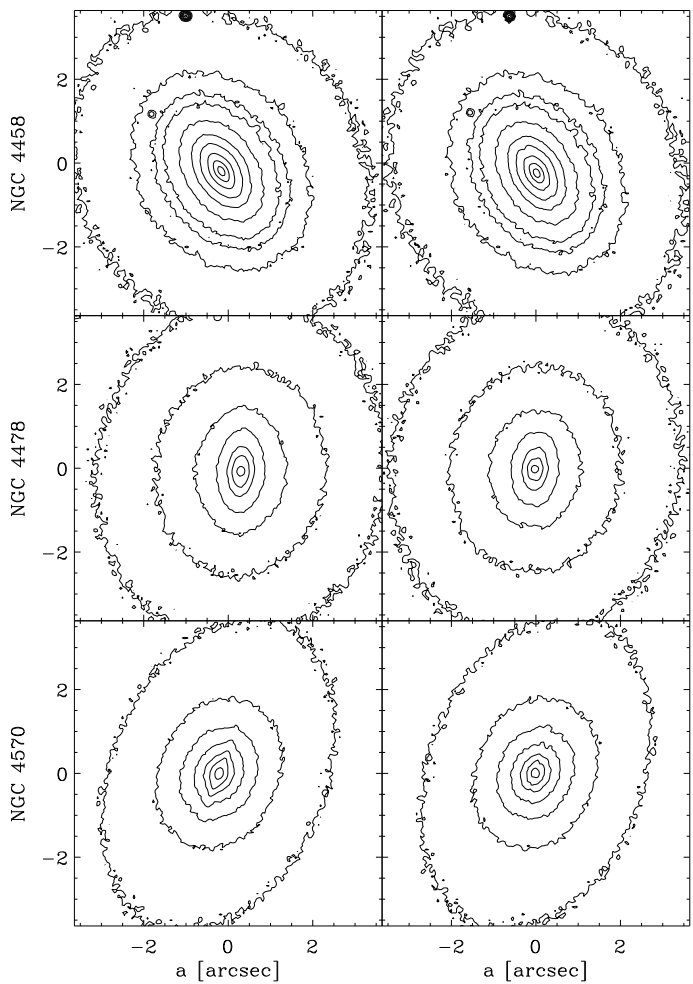}
\caption[]{Contour plots of the WFPC2/F555W deconvolved images of
  NGC 4458, NGC 4478, and NGC 4570 before (left panels) and after
  (right panels) the nuclear disk subtraction. Scales are in
  arcseconds, and orientations are the same as in
  Fig. \ref{fig:unsharp}.}
\label{fig:discsub}
\end{figure}

The $a_4$ Fourier coefficients observed in the three galaxies drop to
zero at small radii ($<0\farcs3$), where the galaxy isophotes become
rounder ($\epsilon < 0.4$) and perfectly elliptical
(Figs. \ref{fig:plot4458} -- \ref{fig:plot4570}). The NSD models have
elliptical isophotes of higher ellipticity ($\epsilon \simeq 0.8$;
Table~\ref{tab:disk_parameters}). When they are subtracted from the
galaxy images, the innermost residual isophotes becomes boxy
($a_4<0$). This is an artefact of the Richardson-Lucy deconvolution
method. As discussed by \citet{mich96}, the algorithm fails to recover
the geometrical properties of the galaxy cores. In particular, the
restored ellipticity is underestimated and decreases to zero towards
the center \citep[see also][]{lucy94}. \citet{vdbetal98} place a lower
limit of $0\farcs3$ for recovering reliable geometric parameters in
HST/WFPC2 images, and \citet{scovdb98} adopted this limit in their
nuclear disk decomposition of NGC~4342 and NGC~4570. We assume that
the results obtained by \citet{vdbetal98} are directly applicable to
our case: because we study images obtained with similar or longer
integration times, galaxies with shallower surface-brightness
profiles and NSDs with equal or larger scale lengths. The
decomposition results do not depend on the surface-brightness
distribution in this region.
Figure~\ref{fig:discsub} shows the shape of the isophotes of the sample
galaxies before and after the subtraction of the best-fit model of
the NSD. The WFPC2/F555W images are considered here as an example. The
diskiness of the galaxy isophotes disappears after the model
subtraction. The same behaviour is also observed in the other
available images.

\begin{figure}
\includegraphics[height=9cm,angle=90]{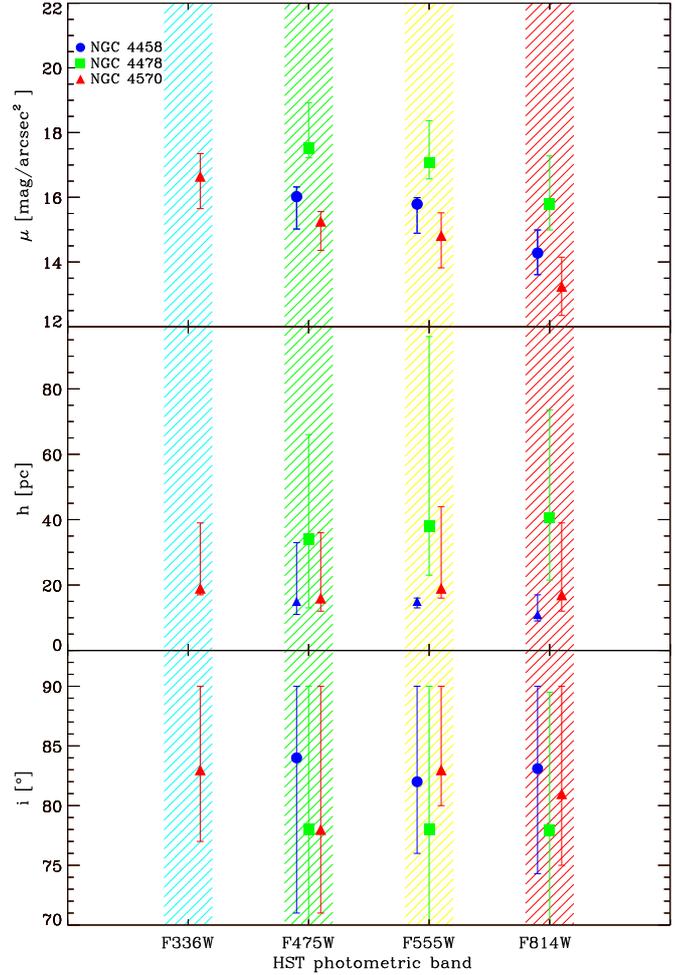}
\caption[]{Photometric parameters of the nuclear stellar disks of
  NGC~4458 ({\em blue circles}), NGC~4478 ({\em green squares}), and
  NGC~4570 ({\em red triangles}) in the WFPC2/F336W, ACS/F475W,
  WFPC2/F555W, and WFPC2/F814W passbands.
}
\label{fig:parameters}
\end{figure}

\smallskip
\noindent
{\em NGC~4458.} 
To find the best-fit model of NGC~4458, we adjusted the disk
parameters until the galaxy isophotes in the inner $1''$ have $a_4
\simeq 0.01$ instead of being perfectly elliptical
(Fig. \ref{fig:plot4458}).
The scale length, inclination, and position angle of the nuclear disk
measured in the ACS/F475W and WFPC2/F555W passbands are consistent
within the errors. They are also consistent within the errors with the
values measured by \citet{moreetal04} in the WFPC2/F814W image
(Fig.~\ref{fig:parameters}). The mean values of the parameters are
$\langle h \rangle = 13$ pc, $\langle i \rangle = 83^\circ$, and $\rm
\langle PA \rangle = 7^\circ$, respectively. It is the smallest and
least luminous NSD ($L_{\rm F814W} = 3 \cdot 10^6$ \lsun) of the
sample. Its light contribution peaks at $0\farcs3$, where $D/T$ ranges
from about 0.4 (ACS/F475W) to 0.2 (WFPC2/F555W) depending on the
passband (Fig. \ref{fig:plot4458}).

\smallskip
\noindent
{\em NGC~4478.} 
The scale length, inclination, and position angle of the nuclear disk
in the ACS/F475W and WFPC2/F555W passbands are consistent within the
errors with the WFPC2/F814W values by
\citet[][Fig.~\ref{fig:parameters}]{moreetal04}. The mean values of
the parameters are $\langle h \rangle = 37$ pc, $\langle i \rangle =
78^\circ$, and $\rm \langle PA \rangle = 153^\circ$, respectively. The
nuclear disk contributes a maximum $D/T=0.2$ in the radial range
between $0\farcs2$ and $0\farcs6$ from the center
(Fig. \ref{fig:plot4478}). It is one of the largest and more luminous
NSDs ($L_{\rm F814W} = 2\cdot10^7$ \lsun) known to date.

\smallskip
\noindent
{\em NGC~4570.}
To find the best-fit model of NGC~4570, we adjusted the disk
parameters until the galaxy isophotes in the inner $1''$ have $a_4
\simeq -0.01$ instead of being perfectly elliptical
(Fig. \ref{fig:plot4570}). In this way, we account for the isophotal
shape of the galaxy bulge, which is characterized by an increasing
boxiness to $a_4 \simeq -0.01$ going from $3''$ to $1''$.
The central surface-brightness, scale length, and inclination of the
NSD were already obtained from the analysis of the WFPC2/F555W image
by \citet{vdb98}. We also decided to perform the photometric
decomposition using our method too. Homogeneous decompositions
and errors in the fitted parameters are required to properly compare
the structural parameters of the NSD in the different passbands. Our
results agree within the errors with those by \citet{vdb98}.
Moreover, the averaged scale length, inclination, and position angle
remain constant within the errors in all the observed passbands
(Fig.~\ref{fig:parameters}). Their mean values are $\langle h \rangle
= 18$ pc, $\langle i \rangle = 81^\circ$, and $\rm \langle PA \rangle
= 158^\circ$, respectively. The disk luminosity ranges from $5 \times
10^6$ (WFPC2/F336W) to $2 \times 10^7$ \lsun\ (WFPC2/F814W). Its
maximum $D/T \simeq 0.5$ is observed at about $0\farcs3$
(Fig. \ref{fig:plot4570}).

The trend of the central surface brightness in the different passbands
is similar for all the measured NSDs. The value of the central surface
brightness is constant within the errors in the ACS/F475W and
WFPC2/F555W passbands and decreases by $1.6$ mag~arcsec$^{-2}$ in the
WFPC2/F814W band. The WFPC2/F336W image was available only for
NGC~4570. The central value of the surface brightness in this filter
is about $1.3$ mag~arcsec$^{-2}$ greater than that in ACS/F475W
passband. In Table \ref{tab:color}, we report the colors NSDs and of
the their host spheroids. The colors of the spheroid were measured
along the disk major axis at a distance of 10 scale lengths on both
sides of the center, where the contribution of the NSD to the total
light is negligible (Figs. \ref{fig:plot4458} -- \ref{fig:plot4570}).
The conversion to the Johnson system was calculated using SYNPHOT
in IRAF.  Since this correction depends on the spectral energy
distribution of the object, it was calculated using the
\citet{kineetal96} spectral templates. The resulting colors are
similar to those measured in galaxies hosting a NSD (NGC~4128,
NGC~4621, and NGC~5308, \citealt{krajaf04}; NGC~4486A,
\citealt{kormetal05}.)
The large uncertainties in the central surface brightnesses are
intrinsic to the photometric decomposition method. This prevented us
from performing color analysis as a consistency check with the
measurement of age and metallicity performed by \citet{vdbetal98},
\citet{moreetal04}, and \citet{krajaf04} and based on the
line-strength indices.

\begin{table}
\caption{Colors of the NSDs and their host spheroids}
\label{tab:color}
\begin{center}
\begin{tabular}{@{}lccccccccc@{}}
\hline
\hline
\noalign{\smallskip}
Color & \multicolumn{2}{c}{NGC~4458} & 
  \multicolumn{2}{c}{NGC~4478} & \multicolumn{2}{c}{NGC~4570}\\
      &  NSD & Spheroid & NSD & Spheroid & NSD & Spheroid \\
\hline
\noalign{\smallskip}
$U-B$ &   ...  &  ...  & ...   & ... & 1.1 & 0.8  \\
$B-V$ &   0.6  &  1.0  & 0.8   & 1.1 & 0.9 & 1.0  \\
$V-I$ &   1.4  &  1.1  & 1.2   & 1.1 & 1.5 & 1.2  \\
 \noalign{\smallskip}
   \hline
\end{tabular}
\end{center}
\begin{minipage}{9 cm}
\begin{small}
{NOTES: The conversion to the Johnson system has been calculated
  as $U-B = [{\rm F336W-F475W}]-0.21$; $B-V = [{\rm
      F475W-F555W}]+0.36$; $V-I = [{\rm F555W-F814W}]-0.05$.}
\end{small}
\end{minipage}
\end{table}

The scale length, inclination, and position angle derived for each NSD
in the different filters are constant within the errors, independently
of their values and the properties of their host galaxy. The
nuclear disk has the same position angle as the inner part of the host
spheroid.

\section{Discussion and conclusions}

We have investigated the photometric properties of the NSDs of three
early-type galaxies in the Virgo cluster, NGC~4458, NGC~4478, and
NGC~4570 by analyzing the WFPC2 and ACS images of their
nuclei available in the HST Science Archive.
The images were unsharp masked and their visual inspection suggested
that the size, orientation, and location of each NSD was independent of
the observed passband.
To verify this, we derived the central surface brightness, scale
length, inclination, and position angle of each NSD in all the
passbands by applying the photometric decomposition method of
\citet{scoben95}.  Some of the images had already been analyzed by
\citet[][ WFPC2/F555W for NGC~4570]{vdbetal98} and
\citet[][WFPC2/F814W for NGC~4458 and NGC~4478]{moreetal04}.  Since we
adopted an identical procedure to that of \citet{moreetal04}, we
relied on their WFPC2/F814W decomposition of NGC~4458 and NGC~4478 and
repeated the WFPC2/F555W analysis of NGC~4570 to build homogeneous
set of photometric parameters and corresponding errors for all the
NSDs in all the passbands.

The structural parameters (i.e., scale length, inclination, and
position angle) of each NSD are constant within the errors in all the
observed passbands, independently of their values and the properties
of their host spheroid. This implies an absence of color gradients in
nuclear disks and can be used to constrain their star formation processes.
NSDs are considered to be the end result of the formation of stars
from gas of external (i.e., accreted from galaxy neighborhoods or
captured during a merger with a gas rich-companion; e.g., NGC~4458,
NGC~4478 \citealt{moreetal04}) or internal origin (i.e., conveyed from
outer galactic regions by a bar; e.g., NGC~4570 \citealt{vdbems98})
piled up in the galaxy nucleus.
Thus independent of the way in which the gas was originally
accumulated towards the center, the dissipational formation process
has to be invoked to account for the dynamically cold structure of the
nuclear disks. This is hard to explain if NSDs assembled from already
formed stars.
We interpret the absence of color gradients in NDSs as the signature
of a star formation event that occurred homogeneously over the entire
disk. Although our sample is admittedly very small, an inside-out
formation scenario \citep{munoetal07} seems to be ruled out in NSDs
given that this would produce radial population gradients that are not
observed.

\begin{acknowledgements}

We would like to thank Enzo Brocato, Valentin Ivanov, Jairo
M\'endez-Abreu and Lodovico Coccato for their useful suggestions.
LM is supported by grant CPDR061795/06 from Padua University.
MC gratefully acknowledges the European Southern Observatory (ESO) for
support via ESO Studentship program at the ESO Research Facilities in
Santiago.
This research has been made possible by support from Padua University
through grants CPDA068415/06, CPDA089220/08, and CPDR095001/09 and from Istituto
Nazionale di Astrofisica (INAF) through grant PRIN2005/32.
This research has made use of the Lyon Extragalactic Database (LEDA)
and NASA/IPAC Extragalactic Database (NED).

\end{acknowledgements}


\bibliographystyle{aa} 
\bibliography{loreb} 

\end{document}